\documentstyle[12pt]{article}
\def\sp{\phantom{a}}

\makeatletter
\@addtoreset{equation}{section}

\makeatother

\begin{document}
\sloppy
\sloppy
\sloppy

\begin{flushright}{ UT-861}
\end{flushright}

\vskip 2.5 truecm

\begin{center}
{\Large  A configuration of 11-dimensional curved superspace 
         as backgrounds for supermembrane }
\end{center}
\vskip 2 truecm
\centerline{\bf Shibusa Yuuichirou}
\vskip .4 truecm
\centerline {\it Department of Physics,University of Tokyo}
\centerline {\it Hongo 7-3-1,Bunkyo-ku,Tokyo 113-0033,Japan}
\vskip 4 truecm

\begin{abstract}

We calculate all of vielbein superfields up to second order in anticommuting 
coordinates in terms of the component fields of 11-dimensional 
on-shell supergravity by using `Gauge completion'. 
This configuration of superspace holds the $\kappa $-symmetry for 
supermembrane Lagrangian and represents 11-dimensional on-shell supergravity.

\end{abstract}

\newpage

\section{Introduction}
Some years ago, T. Banks, W. Fischler, S. H. Shenker and L. Susskind
(BFSS) proposed
that Matrix theory gives a complete description of light-front 
M-theory~\cite{BFSS}. It had been proposed as a theory of D0-branes by 
E. Witten~\cite{w96}.

Up to now, it has become clear that 
Matrix model encodes a remarkable amount of the structure of M-theory
and 11-dimensional supergravity(for reviews, see~\cite{B98}).
The interaction between gravitons in Matrix theory has been shown to
agree with supergravity to some extent~\cite{oka}.

However, this theory is constructed on flat spacetime, therefore Matrix 
theory on curved backgrounds is required. For single D0-branes, the 
theory on curved backgrounds is expected to be described by 
Born-Infeld action~\cite{lei}. For multi-particle system of D0-branes,
namely Matrix model, the theory on curved backgrounds is as yet unknown.
There are many trials to this problem. For example, starting from flat 
Matrix theory, backgrounds are produced by many D0-branes~\cite{yon}.
The other idea is that it is expected as supermembrane on curved 
backgrounds~\cite{wit98}. In this paper we adopt the later idea.

The theory of supermembrane is described as nonlinear sigma model~\cite{ts87}.
Supermembrane consistently couples to 11-dimensional superspace
backgrounds that satisfy a number of constraints which are equivalent 
to 11-dimensional on-shell supergravity~\cite{kap}. After light cone 
gauge fixing and $\kappa $-symmetry gauge fixing, supermembrane theory 
on flat backgrounds is equivalent to a quantum-mechanical model with 
supersymmetric U(N) gauge symmetry in the large N limit by use of matrix 
regularization~\cite{wit88}. It has a continuous mass spectrum and 
instability~\cite{wit89}, therefore it is expected that supermembrane 
matrix theory describes second quantization of D0-branes~\cite{ts96}.
From the beginning of sigma model, it couples to general backgrounds, 
therefore it is expected that sigma model on curved backgrounds 
is a candidate of Matrix theory on curved backgrounds. Actually 
curved backgrounds for supermembrane were investigated~\cite{wit98}.
In this reference, they calculated part of vielbein superfields and all
of 3-form superfields up to second order in anticommuting coordinates.

On the other hand, there are two more aspects of importance of searching
for the structure of 11-dimensional superspace. One is getting much 
knowledge of many low dimensional supergravity theories which can be 
obtained by dimensional reduction from 11-dimensional supergravity. 
The other aspect is getting the Lagrangian of superparticle coupled to 
11-dimensional curved backgrounds. It is not Matrix theory, but very 
similar to Matrix theory~\cite{hy} . 

However the structure of 11-dimensional superspace was not yet well-known. 
Thus we investigate the more higher components of 11-dimensional
superspace. 

In this paper, we compute all of vielbein superfields up to second 
order in anticommuting coordinates in terms of the component fields 
of 11-dimensional on-shell supergravity by using `Gauge completion'.
This configuration of superspace  holds the $\kappa $-symmetry for 
supermembrane 
Lagrangian and represents 11-dimensional on-shell supergravity.  
 
The paper is organized as follows. In section 2, we explain our
notations of the 11-dimensional 
supergravity and obtain the full algebra of transformations in component 
formalism. In section 3, we review the
supermembrane theory and the condition of $\kappa $-symmetry and
explain our notations of the superspace 
geometry and obtain the full algebra of transformations in superspace 
and solve Bianchi identity in our obtained constraints.  
In section 4, we explain `gauge completion' and compute part of 
the superfields. Other notations and conventions 
used throughout this paper are summarized in Appendix.

\section{11-dimensional supergravity}
Supergravity in 11-dimensional spacetime is based on 
`elfbein' field $e_m^{\sp a}$, a Majorana gravitino field 
$\psi _m^{\sp \alpha }$ and third rank antisymmetric gauge field $C_{klm}$.
Its Lagrangian can be written as follows~\cite{CJ78}\cite{wit98}.
\begin{eqnarray}
 L &=& -\frac{1}{2}eR -2e\bar{\psi }_m \Gamma ^{mnl}D_n(\frac{1}{2}(\omega + 
       \hat{\omega }))\psi _l - \frac{1}{96}eF^2 \nonumber \\
   & & +\frac{1}{41472}\epsilon ^{m_1 ... m_{11}}F_{m_1 ... m_4}
       F_{m_5 ... m_8}C_{m_9 ... m_{11}} \nonumber \\
   & & +\frac{1}{96}e(\bar{\psi }_n \Gamma ^{m_1 ... m_4 nl}\psi _l +
        12\bar{\psi }^{m_1}\Gamma ^{m_2 m_3}\psi ^{m_4})(F+\hat{F})_
        {m_1 ...m_4}.
\end{eqnarray}
where $e = det e_m^{\sp a}$, and $\omega _{m\sp b}^{\sp a}$denotes the spin 
connection
\begin{eqnarray}
 \omega _{m\sp b}^{\sp a} &=& -e^{na}\partial _{[m}e_{n]b}+e^{la}e^n_{\sp b}
                              e_m^{\sp c}\partial _{[l}e_{n]c}+e^n_{\sp b}
                              \partial _{[m}e_{n]}^{\sp a} \nonumber \\
                          & & +(\bar{\psi _m}\Gamma _b \psi ^a
                                 +\bar{\psi _b}\Gamma _m \psi ^a
                                 -\bar{\psi _m}\Gamma ^a \psi _b)-\frac{1}{2}
                                 \bar{\psi _n}\Gamma ^{\sp a \sp np}_
                                 {m \sp b}\psi _p ,
\label{conn}
\end{eqnarray}
and $F_{klmn} (=4\partial _{[k}C_{lmn]})$ denotes the field strength of the antisymmetric tensor.
$R (= e^n_{\sp b} e^m_{\sp a} R_{mn}^{\sp \sp \sp ab})$ denotes 
the  scalar curvature, where $R_{mn \sp b}^{\sp \sp \sp a}(= 2\partial_{[m}
\hat{\omega}_{n]\sp b}^{\sp \sp a}-[\hat{\omega}_m,
\hat{\omega}_n]^a_{\sp b})$ denotes curvature tensor. 

The derivative $D_m\epsilon (\equiv (\partial _m - \frac{1}{4}
\omega _{mab} \Gamma ^{ab})\epsilon)$ is covariant with respect 
to local Lorentz transformations. 

The equations of motion are as follows,
\begin{eqnarray}
\label{eom2}
  R_{mn}(\hat{\omega}) &=& \frac{1}{144}g_{mn}\hat{F}_{abcd}\hat{F}^{abcd}
                           -\frac{1}{12}\hat{F}_{mabc}\hat{F}_n^{\sp abc}, \\
                    0  &=& \Gamma^{mnl}\hat{D}_n(\hat{\omega})\psi_l,  \\
  D_a\hat{F}^{abcd}    &=& -\frac{1}{1152}\epsilon^{bcda_1..a_8}
                           \hat{F}_{a_1..a_4}\hat{F}_{a_5..a_8} ,  
\end{eqnarray}
where $R_{mn}(=e_n^{\sp a}e^{lb}R_{mlab})$ denotes Ricci tensor.

Supersymmetry transformations are equal to
\begin{eqnarray}
\label{st}
  \delta _s e_m ^{\sp a} &=      & 2 \bar{\epsilon }\Gamma ^a \psi _m, 
                                  \nonumber \\
  \delta _s \psi _m      &=      & D_m(\hat{\omega })\epsilon + 
                                  T_m ^{\sp rstu}\epsilon \hat{F}_{rstu} 
                                  \equiv \hat{D}_m(\hat{\omega })
                                  \epsilon, \nonumber \\
  \delta _s C_{klm}      &=      & -6 \bar{\epsilon }\Gamma _{[kl}\psi _{m]},\\
  \mbox{with} ,  
  T_m ^{\sp rstu}        &\equiv & \frac{1}{288}(\Gamma _m^{\sp rstu}-8
                                     \delta _m^{[r}\Gamma ^{stu]}),
\end{eqnarray}
where $\hat{F}(= F_{klmn}+12 \bar{\psi}_{[k}\Gamma _{lm}\psi _{n]})$ is 
the supercovariant field strength,
and $\hat{\omega }(= \omega _{m \sp b}^{\sp a}+\frac{1}{2}\bar{\psi _n}
\Gamma ^{\sp a \sp np}_{m \sp b}\psi _p )$ is the supercovariant spin 
connection. 

Note that the spin connection $\omega $ has supersymmetry variation
according to elfbein and gravitino's variation in 2nd-order 
formalism~\cite{des76}. While in 1.5-order formalism, it is defined
as a dependent field determined by its equation of motion, whereas its 
supersymmetry variation is treated as if it were an independent 
field~\cite{wes77}. In this paper we use 2nd-order formalism.

The gauge transformations are equal to 
\begin{eqnarray}
\label{ct}
  \delta _c C_{mnl} = 3 \partial _{[m} \xi_{nl]}.
\end{eqnarray}
The local Lorentz transformations are equal to
\begin{eqnarray}
\label{lt}
  \delta _l e_m^{\sp a}  &=& \lambda ^a_{\sp b}e_m ^{\sp b}, \nonumber \\
  \delta _l \psi 
   _m^{\sp \alpha}       &=& \frac{1}{4}\lambda_{ab}\Gamma^{ab\alpha }_{
                              \sp \sp \sp \beta}\psi _m ^{\sp \beta },
                              \nonumber \\
  \delta_l \omega 
   ^{\sp a}_{m\sp b}     &=& \partial _m \lambda^a_{\sp b}+\lambda^a_
                              {\sp c}\omega _{m\sp b}^{\sp c}- \omega 
                             _{m\sp c}^{\sp a}\lambda ^c_{\sp b}.
\end{eqnarray}
The general coordinate transformation are equal to 
\begin{eqnarray}
\label{gt}
 \delta _g e_m^{\sp a}    &=& \xi^n\partial _n e_m^{\sp a}+\partial _m
                              \xi ^n e_n ^{\sp a}, \nonumber \\
 \delta _g\omega _{m
   \sp b}^{\sp a}         &=& \xi^n\partial _n \omega _{m\sp b}^{\sp a}
                              +\partial _m\xi ^n \omega _{n\sp b}^{
                              \sp a}, \nonumber \\
 \delta _g \psi _m^
  {\sp a}                 &=& \xi^n\partial _n \psi _m^{\sp a}+\partial _m
                              \xi ^n \psi _n ^{\sp a}, \nonumber \\
 \delta _g C_{mnl}        &=& \xi^k\partial _k C_{mnl}+3\partial _{[m}
                              \xi ^k C_{|k|nl]}.
\end{eqnarray}

We obtain the full algebra of these transformations as follows
\begin{eqnarray}
 [\delta _g(\xi _1)+\delta _s(\epsilon _1)+\delta _l(\lambda _1)+\delta _c
    (\xi_{1mn}),\delta _g(\xi _2)+\delta _s(\epsilon _2)+\delta _l(\lambda 
   _2)+\delta _c(\xi_{2mn})]  \nonumber \\
 = \delta _g(\xi _3)+\delta _s(\epsilon _3)+
    \delta _l(\lambda _3)+\delta _c(\xi_{3mn}),
\end{eqnarray}
where 
\begin{eqnarray}
\label{alg}
 \xi _3^m     &=& \xi _2^n \partial _n \xi _1^m + \bar{\epsilon }_2\Gamma ^m
                  \epsilon _1 -(1 \leftrightarrow 2), \nonumber \\
 \epsilon _3  &=& -\bar{\epsilon }_2\Gamma ^n\epsilon _1 \psi _n - \xi _1^n
                  \partial _n \epsilon_2 + \frac{1}{4}\lambda_{2cd}\Gamma^{
                   cd}\epsilon _1 -(1 \leftrightarrow 2), \nonumber \\
 \lambda _
 {3\sp b}^{
  \sp a}     &=& -\bar{\epsilon }_2\Gamma ^n\epsilon _1 \hat{\omega }_{n\sp 
                 b}^{\sp a}-\xi _1^n \partial _n\lambda _{2\sp b}^{\sp a} +
                 \lambda _{2\sp c}^{\sp a}\lambda _{1\sp b}^{\sp c} \nonumber 
                 \\
             & & +\frac{1}{144}\bar{\epsilon }_2(\Gamma _{\sp b}^{a\sp rstu}
                 \hat{F}_{rstu}+24\Gamma _{rs}\hat{F}_{\sp b}^{a \sp rs})
                 \epsilon_1 -(1 \leftrightarrow 2), \nonumber \\
 \xi _{3mn}  &=& -\bar{\epsilon }_2\Gamma ^k\epsilon _1 C_{kmn}
                 -\bar{\epsilon }_2\Gamma _{mn}\epsilon _1-\xi _1^k \partial 
                 _k \xi _{2mn}-2\xi _1^k\partial_{[m}\xi _{2n]k} \nonumber \\
             & & -(1 \leftrightarrow 2). 
\end{eqnarray}

\section{Superspace representation}
\subsection{Supermembrane theory}
Supermembrane theory is described as nonlinear sigma model~\cite{ts87}. It is
written in terms of superspace embedding coordinates 
$Z^M (\xi ) = (X^m (\xi ) , \theta (\xi ))$ , which are functions of the
three world-volume coordinate $\xi ^i (i=0,1,2)$ .

The action is
\begin{eqnarray}
I = \int d^3 \xi (-\frac{1}{2}\sqrt{-g} g^{ij} \Pi _i ^{\sp a}\Pi _j ^{\sp b}
     \eta_{ab} + \frac{1}{2}\sqrt{-g} - \frac{1}{6} \epsilon ^{ijk}
     \Pi _i ^{\sp A} \Pi _j ^{\sp B} \Pi _k ^{\sp C} B_{CBA}), 
\label{mem} 
\end{eqnarray}
where $g_{ij}$ is the metric of the world-volume, $g=det(g_{ij})$ and
$\Pi _i ^{\sp A} \equiv \partial _i Z^M E_M^{\sp A}$.
$E_M^{\sp A}$ is supervielbein, and the 3-form
$B = \frac{1}{6}E^A E^B E^C B_{CBA}$ is potential for the closed 
4-form $H=dB$ .

This action has the following symmetries,

{\it world-volume reparametrization} $\eta ^i(\xi )$
\begin{eqnarray}
  \delta Z^M    &=& \eta ^i \partial _i Z^M , \nonumber \\
  \delta g_{ij} &=& \eta^k \partial _k g_{ij} + 2\partial _{(i}\eta ^k 
                  g_{j)k},
\end{eqnarray}
{\it \sp \sp \sp $\kappa $-symmetry $\kappa ^{\alpha }(\xi )$}
\begin{eqnarray}
  \delta Z^M E_M ^{\sp a}       &=& 0,  \nonumber \\
  \delta Z^M E_M ^{\sp \alpha}  &=& (1+\Gamma ^{\alpha }_{\sp \beta })
                                    \kappa ^{\beta }, \nonumber \\
  \delta (\sqrt{-g}g^{ij})      &=& -2(1+\Gamma ^{\alpha }_{\sp \beta})
                                    \kappa ^{\beta }\Gamma_{ab \sp \alpha
                                      \gamma }\Pi _n^{\sp \gamma }g^{n(i}
                                      \epsilon ^{j)kl}\Pi_k^{\sp a}
                                      \Pi _l^{\sp b} \nonumber \\
                                & & +\frac{-2}{3\sqrt{-g}}
                                      \kappa ^{\alpha }\Gamma _{c \sp 
                                      \alpha \beta }\Pi ^{k \beta }\Pi _k
                                      ^c \epsilon ^{mn(i} \epsilon ^{j)
                                      pq} \nonumber \\
                                & & (\Pi _m^{\sp a}\Pi _{pa}\Pi _n^{
                                      \sp b}\Pi _{qb} +\Pi_m^{\sp a}\Pi _{pa}
                                      g_{nq}+g_{mp}g_{nq}),
\end{eqnarray}
where $\kappa ^{\alpha }(\xi )$ is anticommuting space time spinor and 
the matrix 
$\Gamma $ is defined by
\begin{eqnarray}
 \Gamma = \frac{1}{6 \sqrt{-g}}\epsilon ^{ijk}\Pi _i^{\sp a}
          \Pi _j^{\sp b} \Pi _k^{\sp c}\Gamma _{abc}.  
\end{eqnarray}
Up to surface terms the $\kappa $-invariance of this action imposes the 
following constraints on the 11-dimensional superspace 
geometry~\cite{kap}.
\begin{eqnarray}
\label{const}
 T^a_{\sp \alpha \beta}            &=& -2 \Gamma ^a _{\sp \sp 
                                       \alpha \beta}, \nonumber \\
 H_{\alpha \beta\ ab}              &=& 2 \Gamma _{ab \sp 
                                       \alpha \beta}, \nonumber \\
 H_{\alpha \beta \gamma \delta }=
 H_{\alpha \beta \gamma d}=
 H_{\alpha bcd}                    &=& 0, \nonumber \\
 T^{\alpha }_{\sp \beta \gamma}=
 T^a_{\sp bc}=T^a_{\sp b \gamma }  &=& 0.
\label{kapcon}
\end{eqnarray}

If we want interaction terms up to n-th order in anticommuting
coordinates in Matrix theory, $E_m^a$ and $B_{mnl}$ are required up to
n-th order in anticommuting coordinates, $E_{\mu}^a$ and $B_{\mu mn}$
are required up to (n-1)-th order in anticommuting coordinates,
$B_{\mu \nu m}$ is required up to (n-1)-th order in anticommuting
coordinates, $B_{\mu \nu \rho}$ is required up to (n-2)-th order in 
anticommuting coordinates.

\subsection{Superspace formalism}
In this subsection, we explain notations of the superspace geometry and 
obtain the full algebra of transformations in 11-dimensional superspace. 
As usual, we suppose
that the 11-dimensional superspace has Lorentzian tangent space structure and 
the vielbein $E_M^{\sp A}$ and the connection
 $\Omega _{MA}^{\sp \sp \sp B}$ and 3-form potential $B$ and its field 
strength $H$. 

The Lorentzian assumption implies 
\begin{eqnarray}
 \Omega _{ab}            &=& -\Omega _{ba},  \nonumber \\
 \Omega _{\alpha b}      &=& 0,              \nonumber \\
 \Omega _{\alpha \beta } &=& \frac{1}{4}\Omega _{ab}\Gamma ^{ab}_{\sp \sp \sp
                              \alpha \beta }.
\end{eqnarray}
From these basic fields we can define the torsion ($T^A \equiv DE^A
 = dE^A + E^B \Omega_B^{\sp A}$) and curvature ($R_A^{\sp B} \equiv 
d\Omega _A ^{\sp B}+ \Omega _A^{\sp C}\Omega _C^{\sp B}$). 
Bianchi identity under constraints (\ref{const}) is as follows,
\begin{eqnarray}
\label{bian}
  R_{adb}^{\sp \sp \sp d} &=& \frac{1}{144}\eta_{ab}H_{cdef}
                              H^{cdef}
                           -\frac{1}{12}H_{acde}H_b^{\sp cde}, \\
                    0     &=& \Gamma^{a\alpha}_{\sp \sp \sp \beta}T_{ab}^{\sp                               \sp \sp \beta}, \\
  D_a H^{abcd}       &=& -\frac{1}{1152}\epsilon^{bcda_1..a_8}
                              H_{a_1..a_4}H_{a_5..a_8},  \\
  D_{[a}H_{bcde]}         &=&  0,   \\
  R_{[abc]d}= R_{a[bcd]}  &=&  0,   \\
  R_{\alpha \beta ab}     &=& -\frac{1}{3}H_{abcd}\Gamma^{cd}_{\sp \sp 
                              \alpha \beta}-\frac{1}{72}H_{cdef}\Gamma_{ab
                              \sp \sp \sp \sp \alpha \beta}^{\sp 
                              \sp cdef},   \\
  R_{\beta dca}           &=& T_{cd}^{\sp \sp \epsilon}\Gamma_{a 
                              \epsilon \beta}+T_{da}^{\sp \sp \epsilon}
                              \Gamma_{c \epsilon \beta}-
                              T_{ac}^{\sp \sp \epsilon}\Gamma_{d 
                              \epsilon \beta}, \\ 
 T_{a\sp\beta}^{\sp\alpha}&=& \frac{1}{36}H_{abcd}\Gamma^{bcd\alpha}_{\sp \sp 
                              \sp \sp \beta}-\frac{1}{288}H_{bcde}\Gamma^{
                              \sp bcde \alpha}_{a\sp \sp \sp \sp \sp \beta}\\
 D^{\alpha}H_{abcd}       &=& -12 \Gamma_{[ab \sp \beta}^{\sp \sp \alpha}
                               T_{cd]}^{\sp \sp \beta} 
                              -\frac{1}{7}(\Gamma_{[ab}\Gamma^{ef}DH_{cd]ef})
                              ^{\alpha} .   
\end{eqnarray}
The supertransformation is equal to 
\begin{eqnarray}
\label{TT}
 \delta _T X_{M_p ... M_1} = \Xi ^K \partial _K X_{M_p ... M_1}+p\partial 
                             _{[M_p} \Xi ^K X_{|K|M_{p-1}...M_1]}
\end{eqnarray}
for p-form's components.
The local Lorentz transformations are equal to 
\begin{eqnarray}
\label{LT}
 \delta _L E^A              &=& E^B \Lambda _B^{\sp A},  \nonumber \\
 \delta _L \Omega_B^{\sp A} &=& -\Lambda_B^{\sp C}\Omega _C^{\sp A} +
                                 \Omega _B^{\sp C}\Lambda _C^{\sp A}-
                                d\Lambda _B ^{\sp A} .
\end{eqnarray}
The supergauge transformations are equal to 
\begin{eqnarray}
\label{GT}
  \delta _G B_{LMN} = 3\partial _{[L}\Xi _{MN]} .
\end{eqnarray}
We obtain the full algebra of these transformations as follows
\begin{eqnarray}
 [\delta _T(\Xi _1)+\delta _L(\Lambda _1)+\delta _G
    (\Xi_{1MN}),\delta _T(\Xi _2)+\delta _L(\Lambda 
   _2)+\delta _G(\Xi_{2MN})]  \nonumber \\
    = \delta _T(\Xi _3)+
    \delta _L(\Lambda _3)+\delta _G(\Xi_{3MN}),
\end{eqnarray}
where,
\begin{eqnarray}
\label{ALG}
 \Xi _3^K                   &=& \Xi _2^L\partial _L \Xi _1^K +\delta _1 
                                \Xi _2^K-(1 \leftrightarrow 2), \nonumber \\
 \Lambda _{3A}^{\sp \sp B}  &=& -\Xi _1^K \partial _K\Lambda _{2A}^
                                {\sp \sp B} +\delta _1\Lambda _{2A}^
                                {\sp \sp B} +\Lambda_{1A}^{\sp \sp C}
                                \Lambda _{2C}^{\sp \sp B}
                                 -(1 \leftrightarrow 2), \nonumber \\
 \Xi _{3MN}                 &=& \delta_1\Xi _{2MN}-\Xi _1^K\partial _K
                                \Xi _{2MN} -2\partial _{[M}\Xi _{2N]K}
                                \Xi _1^K -(1 \leftrightarrow 2).          
\end{eqnarray}

There are a great number of component fields in superspace. Thus if we try
to identify superspace representation as ordinary supergravity,
there are a great number of unknown degrees of freedom. The method of this
identification is known as `gauge completion'~\cite{gc}. We shall  
explain it in the next section.

\section{Gauge completion}
`Gauge completion' was introduced to identify superspace 
representation as on-shell supergravity~\cite{gc}. In this section we
review this method and calculate part of components of the superfield 
in terms of the on-shell supergravity fields.

Using this method, up to first order in 
anticommuting coordinates, the superfield components was investigated 
by E. Cremmer and S. Ferrara~\cite{kap}.   
Part of components at second order in anticommuting coordinates was 
investigated by B. de Wit, K. Peeters and J. Plefka~\cite{wit98}.

\subsection{Gauge completion procedure}
`Gauge completion' is searching for structures of the superfields and 
superparameters which are compatible with ordinary supergravity. 
That is to say, supertransformations (\ref{TT}) - (\ref{GT}) are
identified as transformations in 11-dimensional spacetime (\ref{st}),(\ref{ct}),
(\ref{lt}),(\ref{gt}) and 
the $\theta =0$ components of superfields and superparameters are 
identified as the fields and parameters of ordinary supergravity.

Firstly, we choose the input data as follows
\begin{eqnarray}
 E_m^{\sp a(0)}         &=& e_m^{\sp a}, \nonumber \\
 E_m^{\sp \alpha (0)}   &=& \psi _m^{\sp \alpha }, \nonumber \\
 \Omega _{mb}^
 {\sp \sp \sp a(0)}       &=& -\hat{\omega }_{m \sp b}^{\sp a}, \nonumber \\
 \Xi ^{m(0)}            &=& \xi ^m, \nonumber \\
 \Xi ^{\mu (0)}      &=& \epsilon ^{\mu }, \nonumber \\
 \Xi ^{(0)}_{mn}        &=& \xi _{mn}, \nonumber \\
 B_{mnl}^{(0)}          &=& C_{mnl}.  
\end{eqnarray}
From (\ref{LT}) and (\ref{lt}), we obtain 
\begin{eqnarray}
 \Lambda _b ^{\sp a \sp (0)} = \lambda _{\sp b}^a .
\end{eqnarray}
Moreover we introduce the assumption that superparameters do not include the 
derivative of $\epsilon$. 
Then, the higher order components in anticommuting coordinates can be
obtained  by requiring consistency between the algebra of 
superspace supergravity and that of ordinary supergravity.

If we can represent $\Xi_{MN} = 2\partial_{[M}\Phi_{N]}$, we can 
choose the gauge as $\Xi_{MN}=0$ because this superparameters do not 
change the 3-form 
superfields (\ref{GT}) and the algebra (\ref{ALG}). 
Thus we can choose the gauge as follows,
\begin{eqnarray}
 \Xi _{\mu N}^{(0)}    &=& 0 .
\end{eqnarray}

To obtain the higher order components of superparameters  
which depend on $\epsilon$, we must calculate the commutation of 
two supersymmetry transformation.

According to (\ref{alg}),(\ref{ALG}) and (\ref{sam}),
\begin{eqnarray} 
[\delta_{s1},\delta_{s2}]E_m^{\sp a (0)}&=& (\Xi_3 ^K\partial _KE_m^{\sp a}
                                            +\partial _m \Xi_3 ^K E_K^{\sp a}
                                            + E_m^{\sp b}\Lambda_{3b}^{\sp \sp
                                            a})|_{\theta =0} \nonumber \\
                                        &=& (\delta_g(2\bar{\epsilon }_2
                                            \Gamma ^m\epsilon _1)+\delta_s(
                                            -2\bar{\epsilon }_2\Gamma ^n
                                            \epsilon _1 \psi _n)+\delta_c(
                                            -2\bar{\epsilon }_2\Gamma ^k
                                            \epsilon _1 C_{kmn}-2\bar{
                                            \epsilon }_2\Gamma _{mn}
                                            \epsilon _1) \nonumber \\
                                        & & +\delta_l(
                                            -2\bar{\epsilon }_2\Gamma ^n
                                            \epsilon _1 \hat{\omega }_{n
                                            \sp b}^{\sp a}+\frac{1}{72}\bar{
                                            \epsilon }_2(\Gamma _{\sp b}^{a
                                            \sp rstu}\hat{F}_{rstu}+24
                                            \Gamma _{rs}\hat{F}_{\sp b}^{a 
                                            \sp rs})\epsilon_1 ))e_m^{\sp a}.
\end{eqnarray}
Thus one obtains 
\begin{eqnarray}
    \Xi^{k(1)}(susy) = \bar{\theta} \Gamma^k \epsilon.
\end{eqnarray}

In the same way, to obtain the higher order components of superparameters  
which depend on $\lambda$ we must calculate the commutation of 
supersymmetry transformation and Lorentz transformation. 
To obtain the higher order components of superparameters 
which depend on $\xi_{mn}$ we must calculate the commutation of 
supersymmetry transformation and gauge transformation.  
To obtain the higher order components of superparameters which depend 
on $\xi^m$ we must calculate the commutation of supersymmetry transformation
and general coordinate transformation. 

By this procedure, the following results had been 
known ~\cite{kap},~\cite{wit98}.
\begin{eqnarray}
 \Xi ^m        &=& \xi ^m +\bar{\theta }\Gamma ^m\epsilon -\bar{\theta }
                   \Gamma ^n\epsilon \bar{\theta }\Gamma ^m\psi _n
                   +{\cal O}(\theta ^3), \\
 \Xi ^{\mu}    &=& \epsilon ^{\mu }-\frac{1}{4}\lambda _{cd}(\Gamma ^{cd}
                   \theta )^{\mu }-\bar{\theta }\Gamma ^n\epsilon 
                   \psi _n ^{\sp \mu }+{\cal O}(\theta ^2), \\
 \Lambda _b^{
  \sp a}       &=& \lambda ^a_{\sp b}-\bar{\theta }\Gamma ^n\epsilon 
                   \hat{\omega }_{n \sp b}^{\sp a}+\frac{1}{144}\bar{\theta }
                   (\Gamma ^{a\sp rstu}_{\sp b}\hat{F}_{rstu}+24\Gamma _{rs}
                   \hat{F}_{\sp b}^{a \sp rs})\epsilon
                   +{\cal O}(\theta ^2), \\
 \Xi _{mn}     &=& \xi _{mn}-(\bar{\theta }\Gamma ^p\epsilon C_{pmn}+
                   \bar{\theta }\Gamma _{mn}\epsilon ) 
                   +\bar{\theta }\Gamma ^k\epsilon \bar{\theta }\Gamma^l
                   \psi _kC_{lmn}+\bar{\theta }\Gamma ^k\epsilon \bar{
                   \theta }\Gamma _{mn}\psi _k +\frac{4}{3}\bar{\theta}
                   \Gamma^l\epsilon\bar{\theta}
                   \Gamma_{l[m}\psi_{n]} \nonumber \\
               & & +\frac{4}{3}\bar{\theta}\Gamma^l\psi_{[n}\bar{\theta}
                   \Gamma_{|l|m]}\epsilon
                   +{\cal O}(\theta ^3), \\
 \Xi _{m\mu }  &=& \frac{1}{6}\bar{\theta}\Gamma^n\epsilon(\bar{\theta}
                   \Gamma_{mn})_{\mu}+\frac{1}{6}(\bar{\theta}\Gamma^n)_{\mu}
                   \bar{\theta}\Gamma_{mn}\epsilon+{\cal O}(\theta ^3), \\
 \Xi _{\mu 
        \nu }  &=& {\cal O}(\theta ^3).
\end{eqnarray}
According to superspace algebra,
\begin{eqnarray}
 \delta_{susy} E_m^{
\sp a}|_{\theta =0}    &=& (\Xi ^K(susy)\partial _KE_m^{\sp a}
                           +\partial _m \Xi^K(susy)E_K^{\sp a}+ E_m^{
                           \sp b}\Lambda 
                           _b^{\sp a}(susy))|_{\theta =0} \nonumber \\
                       &=& \epsilon ^{\nu }\partial _{\nu }
                           (E_m^{\sp a (1)})+\partial_m\epsilon ^{\nu }
                           E_{\nu }^{\sp a(1)},  
\end{eqnarray}
while in ordinary supergravity
\begin{eqnarray}
 \delta_{susy} e_m^{\sp a} = 2\bar{\epsilon }\Gamma ^a \psi _m.
\end{eqnarray}
Thus, one obtains 
\begin{eqnarray}
\label{sam}
 E_{\nu }^{\sp a \sp (0)}  &=& 0, \nonumber \\
 E_m^{\sp a (1)}           &=& 2\bar{\theta }\Gamma ^a\psi _m .
\end{eqnarray}
By this procedure, the following results had been 
known ~\cite{kap},~\cite{wit98}.
\begin{eqnarray}
 E_m^{\sp a}   &=& e_m^{\sp a}+2\bar{\theta }\Gamma ^a\psi _m -\frac{1}{4}
                   \bar{\theta }\Gamma^{acd}\theta \hat{\omega }_{mcd}+
                   \frac{1}{72}\bar{\theta }\Gamma _m^{\sp rst}\theta \hat{F}
                   _{rst}^{\sp \sp \sp a} \nonumber \\
               & & +\frac{1}{288}\bar{\theta }\Gamma ^{rstu}\theta \hat{F}
                   _{rstu}e_m^{\sp a}-\frac{1}{36}\bar{\theta }\Gamma 
                   ^{astu}\theta \hat{F}_{mstu}+{\cal O}(\theta ^3), \\
 E_m^
 {\sp \alpha}  &=& \psi _m^{\sp \alpha}-\frac{1}{4}\hat{\omega }_{mab}(\Gamma 
                   ^{ab}\theta )^{\alpha }+(T_m^{\sp rstu} \theta )^{\alpha}
                   \hat{F}_{rstu}+{\cal O}(\theta ^2), \\
 E_{\mu }^{
 \sp a}        &=&  -(\Gamma ^a \theta)_{\mu }+{\cal O}(\theta ^3),\\
 E_{\mu }^{
 \sp \alpha}   &=& \delta _{\mu }^{\sp \alpha }+{\cal O}(\theta ^2),\\
  \Omega _
 {\mu b}^{
 \sp \sp a}    &=& \frac{1}{144}\{(\Gamma ^{a\sp rstu}_{\sp b}\theta )_{\mu }
                   \hat{F}_{rstu}+24(\Gamma _{rs}\theta )_{\mu }\hat{F}_
                   {\sp b}^{a \sp rs} \} +{\cal O}(\theta ^2),  \\
 \Omega _{mab} &=& \hat{\omega }_{mab}+2\bar{\theta }\{ e^n_{\sp a}e^k_
                   {\sp b}(-\Gamma _kD_{[m}\psi _{n]}+\Gamma _nD_{[m}
                   \psi _{k]}+\Gamma _mD_{[n}\psi _{k]})\} \nonumber \\
               & & +\frac{1}{72}\bar{\theta}(\Gamma_{ab}^{\sp \sp rstu}
                   \hat{F}_{rstu}+24\Gamma_{rs}\hat{F}_{ab}^{\sp \sp rs})
                   \psi_m +{\cal O}(\theta ^2),  \\
 B_{mnl}       &=& C_{mnl}-6\bar{\theta }\Gamma _{[mn}\psi _{l]}+\frac{3}{4}
                   \hat{\omega }_{[l}^{\sp \sp cd}\bar{\theta }\Gamma _{
                   mn]cd}\theta -\frac{3}{2}\hat{\omega }_{[lmn]}\theta ^2
                   \nonumber \\
               & & -\frac{1}{96}\bar{\theta }\Gamma _{mnl}^{\sp \sp \sp rstu}
                   \theta \hat{F}_{rstu}-\frac{3}{8}\bar{\theta }\Gamma _{
                   [l}^{\sp \sp rs}\theta \hat{F}_{|rs|mn]}-12\bar{\theta }
                   \Gamma _a\psi _{[m}\bar{\theta }\Gamma ^a_{\sp n}\psi _{l]}
                   \nonumber \\
               & & +{\cal O}(\theta ^3), \\
 B_{mn\mu }    &=& (\bar{\theta }\Gamma _{mn})_{\mu }+\frac{8}{3}\bar{\theta}
                   \Gamma^k\psi_{[m}(\bar{\theta}\Gamma_{|k|n]})_{\mu}+
                   \frac{4}{3}(\bar{\theta}\Gamma^k)_{\mu}\bar{\theta}
                   \Gamma_{k[m}\psi_{n]}+{\cal O}(\theta ^3), \\
 B_{m\mu \nu } &=& (\bar{\theta }\Gamma _{mn})_{(\mu }(\bar{\theta }\Gamma ^n
                   )_{\nu )}+{\cal O}(\theta ^3), \\
 B_{\mu \nu 
 \rho }        &=& (\bar{\theta }\Gamma _{mn})_{(\mu }(\bar{\theta }\Gamma ^m
                   )_{\nu }(\bar{\theta }\Gamma ^n)_{\rho )}
                   +{\cal O}(\theta ^3). \\
\end{eqnarray}
Because the flat geometry had been known, we include the $\theta ^3$ term 
in $B_{\mu \nu \rho }$ for completeness.

\subsection{Calculation}
$\Xi^{\mu (2)}$ is subject to the following equations,
\begin{eqnarray}
  \epsilon_2^{\nu}
  \partial_{\mu}
  \partial_{\nu} 
  \Xi_1^{\alpha} -(1 
  \leftrightarrow 2)  &=& \frac{1}{576}
                           \epsilon_2^{\nu}(\Gamma_{ab}^{\sp \sp rstu}+
                           24\delta_b^{\sp u}\delta_a^{\sp t}
                          \Gamma^{rs}\epsilon_1)_{\nu}\Gamma^{ab\alpha}_{
                          \sp \sp \sp \mu}\hat{F}_{rstu} \nonumber \\
                      & & +(\Gamma^k\epsilon_1)_{\mu}(T_k^{rstu})_{\sp \nu}^{
                           \alpha}\epsilon_2^{\nu}\hat{F}_{rstu}-\epsilon_2^{\nu}
                           (\Gamma^n\epsilon_1)_{\nu}(\Gamma^k\psi_n)_{\mu}
                           \psi_k^{\sp \alpha} \nonumber \\
                      & & -(\Gamma^n\epsilon_1)_{\mu}
                           \epsilon_2^{\nu}(\Gamma^k\psi_n)_{\nu}\psi_k^{\sp \alpha}
                           -\frac{1}{4}(\Gamma^n\epsilon_1)_{\mu}\hat{\omega}_{nab}
                           (\Gamma^{ab})^{\alpha}_{\sp \nu}\epsilon_2^{\nu}\nonumber \\
                      & & -\frac{1}{4}\epsilon_2^{\nu}(\Gamma^n\epsilon_1)_{\nu}
                           \hat{\omega}_{nab}(\Gamma^{ab})^{\alpha}_{\sp \mu}
                           -(1 \leftrightarrow 2) .
\end{eqnarray}
However, if simply we drive the equation,
\begin{eqnarray}
  \epsilon_2^{\nu}
  \partial_{\mu}
  \partial_{\nu} 
  \Xi_1^{\alpha}      &=& (\Gamma^k\epsilon_1)_{\mu}(T_k^{rstu})_{\sp \nu}^{
                           \alpha}\epsilon_2^{\nu}\hat{F}_{rstu}-\epsilon_2^{\nu}
                           (\Gamma^n\epsilon_1)_{\nu}(\Gamma^k\psi_n)_{\mu}
                           \psi_k^{\sp \alpha}
                           \nonumber \\
                      & &  +\frac{1}{576}
                          \epsilon_2^{\nu}(\Gamma_{ab}^{\sp \sp rstu}+
                           24\delta_b^{\sp u}\delta_a^{\sp t}
                          \Gamma^{rs}\epsilon_1)_{\nu}\Gamma^{ab\alpha}_{
                          \sp \sp \sp \mu}\hat{F}_{rstu} \nonumber \\
                      & & -(\Gamma^n\epsilon_1)_{\mu}
                           \epsilon_2^{\nu}(\Gamma^k\psi_n)_{\nu}\psi_k^{\sp \alpha}
                           -\frac{1}{4}(\Gamma^n\epsilon_1)_{\mu}\hat{\omega}_{nab}
                           (\Gamma^{ab})^{\alpha}_{\sp \nu}\epsilon_2^{\nu}\nonumber \\
                      & & -\frac{1}{4}\epsilon_2^{\nu}(\Gamma^n\epsilon_1)_{\nu}
                           \hat{\omega}_{nab}(\Gamma^{ab})^{\alpha}_{\sp \mu},
\end{eqnarray}
this equation is inconsistent because $\mu$ and $\nu$ in the left-hand 
side of it are antisymmetric but these in the right-hand side of it are not
antisymmetric. Thus we must add terms in the right-hand side of this equation.
\begin{eqnarray}
  \epsilon_2^{\nu}
  \partial_{\mu}
  \partial_{\nu} 
  \Xi_1^{\alpha}      &=& (\Gamma^k\epsilon_1)_{\mu}(T_k^{rstu}\epsilon_2)^{
                           \alpha}\hat{F}_{rstu} +\frac{1}{576}
                           \bar{\epsilon}_2(\Gamma_{ab}^{\sp \sp rstu}+
                           24\delta_b^{\sp u}\delta_a^{\sp t}
                          \Gamma^{rs})\epsilon_1\Gamma^{ab\alpha}_{
                          \sp \sp \sp \mu}\hat{F}_{rstu} \nonumber \\
                      & & +\frac{\hat{F}_{rstu}}{576}[2\bar{\epsilon}_2
                          \Gamma^{cbu}\epsilon_1\Gamma_{bc}^{\sp \sp rst}
                          -6\bar{\epsilon}_2\Gamma^{ctu}\epsilon_1
                          \Gamma_c^{\sp rs}-3\bar{\epsilon}_2
                          \Gamma^{dctu}\epsilon_1\Gamma_{cd}^{\sp \sp rs}
                          -4\bar{\epsilon}_2\Gamma^{dstu}\epsilon_1
                          \Gamma_d^{\sp r} \nonumber \\
                      & & +8\bar{\epsilon}_2\Gamma^{rstu}
                          \epsilon_1\delta]^{\alpha}_{\sp \mu}-\epsilon_2^{\nu}
                           (\Gamma^n\epsilon_1)_{\nu}(\Gamma^k\psi_n)_{\mu}
                           \psi_k^{\sp \alpha} -\frac{1}{4}\epsilon_2^{\nu}
                          (\Gamma^n\epsilon_1)_{\nu}\hat{\omega}_{nab}
                          (\Gamma^{ab})^{\alpha}_{\sp \mu}\nonumber \\
                      & & -(\Gamma^n\epsilon_1)_{\mu}
                           \epsilon_2^{\nu}(\Gamma^k\psi_n)_{\nu}\psi_k^{\sp \alpha}
                           -\frac{1}{4}(\Gamma^n\epsilon_1)_{\mu}\hat{\omega}_{nab}
                           (\Gamma^{ab})^{\alpha}_{\sp \nu}\epsilon_2^{\nu}.
\end{eqnarray}
Thus we obtain
\begin{eqnarray}
\label{kusi}
 \Xi ^{\mu}    &=& \hat{F}_{abcd}[-\frac{1}{1024}\bar{
                   \theta}\theta(\Gamma^{abcd}\epsilon)^{\mu}
                   +\bar{\theta}\Gamma^{efg}\theta(\frac{5}{18432}(
                   \Gamma_{gfe}^{\sp \sp \sp abcd}\epsilon)^{\mu} 
                   -\frac{7}{4608}\delta_g^{\sp d}(\Gamma_{fe}^{\sp \sp abc}
                   \epsilon)^{\mu} \nonumber \\
               & & -\frac{5}{1536}\delta_g^{\sp d}\delta_f^{
                   \sp c}(\Gamma_e^{\sp ab}\epsilon)^{\mu}+\frac{7}{768}
                   \delta_g^{\sp d} \delta_f^{\sp c}\delta_e^{\sp b}(
                   \Gamma^a\epsilon)^{\mu}) 
                   +\bar{\theta}\Gamma^{efgh}\theta(\frac{5}{73728}(\Gamma_{
                   hgfe}^{\sp \sp \sp \sp abcd}\epsilon)^{\mu}\nonumber \\ 
               & & +\frac{1}{4608}\delta_h^{\sp d}(\Gamma_{gfe}^{\sp \sp \sp 
                   abc}\epsilon)^{\mu}+\frac{1}{1024}\delta_h^{\sp d}
                   \delta_g^{\sp c}(\Gamma_{fe}^{\sp \sp ab}\epsilon)^{\mu} 
                    -\frac{1}{2304}\delta_h^{\sp d} 
                   \delta_g^{\sp c}\delta_f^{\sp b}(\Gamma_e^{\sp a}\epsilon)
                   ^{\mu} \nonumber \\
               & & -\frac{73}{9216}\delta_h^{\sp d}\delta_g^{\sp c}
                   \delta_f^{\sp b}\delta_e^{\sp a}
                   \epsilon^{\mu})]+\bar{\theta}\Gamma^n\epsilon\bar{\theta}
                   \Gamma^k\psi_n\psi_k^{\sp \mu}+\frac{1}{4}\bar{\theta}
                   \Gamma^n\epsilon\hat{\omega}_{nab}(\Gamma^{ab}\theta)^{\mu}. 
\end{eqnarray}
$E_{\mu}^{\sp \alpha(2)}$ is subject to the following equation,
\begin{eqnarray}
\epsilon^{\nu}
\partial_{\nu}
E_{\mu}^{
\sp \alpha(2)}  &=& -\partial_{\mu}\Xi^{\alpha (2)}(\hat{F}dependent \sp terms)
                    -(\Gamma^k \epsilon)_{
                    \mu}(T_k^{\sp rstu}\theta)^{\alpha}
                    \hat{F}_{rstu} \nonumber \\
                & & -\frac{\hat{F}_{rstu}}{576}\bar{\theta}(\Gamma_{ab}^{
                    \sp \sp rstu}+24\delta_b^{\sp u}\delta_a^{
                    \sp t}\Gamma^{rs})\epsilon
                    \Gamma^{ab\alpha}_{\sp \sp \sp \mu}.
\end{eqnarray}
From (\ref{kusi}) we obtain
\begin{eqnarray}
 E_{\mu }^{
 \sp \alpha}   &=& \hat{F}_{abcd}[\bar{\theta}
                   \Gamma^{efg}\theta(\frac{1}{576}\delta_g^{\sp d}
                   \Gamma_{fe \sp \sp \sp \sp \mu}^{\sp \sp abc\alpha}
                   -\frac{293}{55296}\delta_g^{\sp d}\delta_f^{\sp c}
                   \Gamma_{e \sp \sp \sp \mu}^{\sp ab\alpha}) \nonumber \\
               & & +\bar{\theta}\Gamma^{efgh}\theta(-\frac{1}{384}
                   \delta_h^{\sp d}\delta_g^{\sp c}\Gamma_{fe\sp \sp 
                   \sp \mu}^{\sp \sp ab\alpha}-\frac{1}{288}\delta_h^{\sp d}
                   \delta_g^{\sp c}\delta_f^{\sp b}\Gamma_{e\sp \sp \mu}^{
                   \sp a\alpha} \nonumber    \\
               & & +\frac{1}{144}\delta_h^{\sp d}\delta_g^{
                   \sp c}\delta_f^{\sp b}\delta_e^{\sp a}\delta^{\alpha}_{
                   \sp \mu})].
\end{eqnarray}
$E_m^{\sp \alpha(2)}$ is subject to the following equation,
\begin{eqnarray}
\epsilon^{\mu}
\partial_{\mu}
E_m^{\alpha}   &=& \bar{\theta}\Gamma^k \epsilon(T_k^{\sp rstu}\psi_m)^{
                   \alpha}\hat{F}_{rstu}-\frac{\hat{F}_{rstu}}{576}\bar{
                   \theta}(\Gamma_{ab}^{\sp \sp rstu}+24\delta_a^{\sp r}
                   \delta_b^{\sp s}\Gamma^{tu})\epsilon (\Gamma^{ab}\psi_m)^{
                   \alpha} \nonumber \\
               & & -\frac{\hat{F}_{rstu}}{288}\bar{
                   \epsilon}(\Gamma_{ab}^{\sp \sp rstu}+24\delta_a^{\sp r}
                   \delta_b^{\sp s}\Gamma^{tu})\psi_m (\Gamma^{ab}\theta)^{
                   \alpha}+2\bar{\epsilon}\Gamma^k \psi_m(T_k^{\sp rstu}
                   \theta)^{\alpha}\hat{F}_{rstu} \nonumber \\
               & & -2\bar{\epsilon}\Gamma^k 
                   \theta (\hat{D}_{[m}\psi_{n]})^{\alpha}
                   -24(T_m^{\sp rstu}\theta)^{\alpha}\bar{\epsilon}
                   \Gamma_{[rs}\hat{D}_t\psi_{u]}+\bar{\epsilon}
                   \Gamma_b\hat{D}_{[m}\psi_{n]}(\Gamma^{nb}\theta)^{
                   \alpha}  \nonumber \\
               & & -\frac{1}{2}\bar{\epsilon}
                   \Gamma_m\hat{D}_{[l}\psi_{n]}(\Gamma^{ln}\theta)^{\alpha}.
\end{eqnarray}
We obtain
\begin{eqnarray}
 E_m^
 {\sp \alpha}  &=& \bar{\theta}\theta[-\frac{1}{384}(\Gamma^{
                   abcd}\psi_m)^{\alpha}\hat{F}_{abcd}] 
                   +\bar{\theta}\Gamma^{efg}\theta[(\frac{1}{2304}
                   \Gamma_{gfe}^{\sp \sp \sp abcd}+\frac{1}{576}\delta_g^{
                   \sp d}\Gamma_{fe}^{\sp \sp abc} \nonumber \\
               & & -\frac{1}{192}\delta_g^{
                   \sp d}\delta_f^{\sp c}\Gamma_e^{\sp ab} 
                   +\frac{1}{96}\delta_g^{\sp d}\delta_f^{\sp c}\delta_e^{
                   \sp b}\Gamma^a)^{\alpha}_{\sp \beta}\psi_m^{\sp \beta}
                   \hat{F}_{abcd}
                   +\frac{1}{8}(\Gamma_{fe}D_{[m}\psi_{g]})^{
                   \alpha} \nonumber \\
               & & +\frac{1}{4}e_{mg}(D_f\psi_e)^{\alpha}]
                   +\bar{\theta}
                   \Gamma^{efgh}\theta[(\frac{1}{27648}
                   \Gamma_{hgfe}^{\sp \sp \sp \sp abcd}+\frac{1}{1728}
                   \delta_h^{\sp d}\Gamma_{gfe}^{\sp \sp 
                   \sp abc} \nonumber \\
               & & -\frac{1}{384}\delta_h^{\sp d}\delta_g^{\sp c}\Gamma_{fe}^{
                   \sp \sp ab}-\frac{1}{288}\delta_h^{\sp d}\delta_g^{\sp c}
                   \delta_f^{\sp b}\Gamma_e^{\sp a} \nonumber \\
               & & -\frac{1}{1152}\delta_h^{
                   \sp d}\delta_g^{\sp c}\delta_f^{\sp b}\delta_e^{\sp a}
                   \delta)^{\alpha}_{\sp \beta}\psi_m^{\sp \beta}\hat{F}_{
                   abcd}-\frac{1}{24}(\Gamma_{gfe}D_{[m}\psi_{h]})^{
                   \alpha}-
                   \frac{1}{24}(\Gamma_{hgm}D_f\psi_e)^{\alpha}\nonumber \\
               & & +\frac{1}{6}e_{mh}(\Gamma_gD_f\psi_e)^{
                   \alpha}]. 
\end{eqnarray}
Thus, we have obtained all components of vielbein
superfields and 3-form superfields up to second order in anticommuting 
coordinates.

This results are consistent with the all transformations calculation.

\subsection{Bianchi and constraints }
From results of the previous subsection, we obtain the torsion fields and 
field strength fields as follows,
\begin{eqnarray}
 T^{a(0)}_{\sp \alpha \beta}       &=& -2 \Gamma ^a _{\sp \sp 
                                       \alpha \beta}, \\
 H_{\alpha \beta\ ab}^{(0)}        &=& 2 \Gamma _{ab \sp 
                                       \alpha \beta}, \\
 H_{\alpha \beta \gamma 
 \delta}^{(0)}=
 H_{\alpha \beta \gamma d}^{(0)}=
 H_{\alpha bcd}^{(0)}              &=& 0, \\
 T^{\alpha (0)}_{\sp \beta 
 \gamma}=T^{a(0)}_{\sp bc}=
 T^{a(0)}_{\sp b \gamma }          &=& 0, \\
 T_{cb}^{\sp \sp \alpha(0)}        &=& 2e^m_{\sp b}e^n_{\sp c}\hat{D}_{[n}
                                       \psi_{m]}^{\sp \alpha},  \\
 T_{c\sp\beta}^{\sp\alpha (0)}     &=& \frac{1}{36}\hat{F}_{cfgh}\Gamma^{fgh
                                       \alpha}_{\sp \sp \sp \sp \beta}-
                                       \frac{1}{288}\hat{F}_{efgh}\Gamma^{
                                       \sp efgh \alpha}_{c\sp \sp \sp \sp 
                                       \sp \beta}, \\ 
 H_{abcd}^{(0)}                    &=& \hat{F}_{abcd} .
\end{eqnarray}
This results satisfy the $\kappa$-symmetry constraints (\ref{const}). 
By using the Bianchi identity (\ref{bian}), we obtain equations of 
motion in component formalism (\ref{eom2}). In particular, the equation of 
motion for gravitino fields can be obtained by using gauge fixing condition,
\begin{eqnarray}
\label{fix}
 (\hat{D}^m\psi_m)^{\alpha}     &=& 0, \nonumber \\
 (\Gamma ^m \psi _m)^{\alpha}   &=& 0. 
\end{eqnarray}
Thus this configuration can be identified as backgrounds for
supermembrane.

\section{Discussion}
We have obtained $\Xi^{\mu (2)}$,$E_M^{\sp \alpha(2)}$. Up to second order in
anticommuting coordinates, $\Lambda_a^{\sp b(2)}$ and $\Omega_{MA}^{\sp 
\sp B(2)}$ remain. These terms and terms which are required to
obtain terms of Matrix theory which are third order in anticommuting 
coordinates is under considerations. The background field linear coupling to 
flat Matrix in all order of anticommuting coordinates was conjectured 
in ref.~\cite{plef} . 

There is a problem of interest to us. It is a gauge fixing problem. As a
previous section, gravitino fields are subject to gauge fixing
conditions (\ref{fix}). Whether for $e_m^{\sp a},c_{lmn}$ gauge fixing 
conditions are required or not, it is not yet obvious. From the
beginning the supermembrane theory has general coordinates
transformation symmetry, local supersymmetry, local Lorentz symmetry 
and U(1) gauge symmetry, however to holds $\kappa$-symmetry, 
we have the constraints for backgrounds which contain gravitino's gauge 
fixing conditions. I think interpretation of $\kappa$-symmetry must be 
investigated further.

\section*{Acknowledgments}

I would like to thank Y.Matsuo for valuable suggestions, and thank
K. Hosomichi for valuable comments and discussions. 
\appendix
\section*{Appendix}
\section{Indices}

We use Greek indices for spinorial components and Latin indices for
vector components. And we use former alphabet for the tangent space
indices and later for general coordinates indices: $a,b,c,...$ for tangent
vector indices and $k,l,m,...$ for general vector indices,
and  $\alpha ,\beta ,...$ for tangent spinorial indices and 
$\mu , \nu ,...$ for general spinorial indices.

Superspace coordinates $(x^m ,\theta ^{\mu })$ are designated 
$Z^M $ , where later capital Latin alphabet $M,N,..$ are collective 
designations for general coordinate indices. While former capital
Latin alphabet $A,B,..$ are collective designations for tangent
space indices.

\section{p-form superfield}

We introduce p-form superfields as follows,
\begin{eqnarray}
  X               &\equiv & \frac{1}{p!} dz^{M_p}...dz^{M_1} 
                            X_{M_p ... M_1} \nonumber \\
                  &\equiv & \frac{1}{p!} E^{A_p}...E^{A_1} X_{A_p ... A_1}, \\
  X_{A_p ... A_1} &\equiv & \sum_{i=1}^{32} X_{A_p ... A_1}^{\sp \sp \sp (i)}.
\end{eqnarray}
 $X_{A_p ... A_1}^{\sp \sp \sp (i)}$ is component at i-th order in 
anticommuting coordinates.

\section{Convention}

Symmetrization bracket $(\sp \sp )$ and antisymmetrization bracket
$[\sp \sp ]$ is defined as follows,

\begin{eqnarray}
   [M_1 ... M_N]   &=& \frac{1}{N!}( M_1 ... M_N \sp +
                               \mbox{antisymmetric terms} ), \nonumber \\
   (M_1 ... M_N)   &=& \frac{1}{N!}( M_1 ... M_N \sp +
                               \mbox{symmetric terms} ).
\end{eqnarray}
 
\section{Gamma matrices(11-dimensional)}

Since we use the Majorana representation, all components are real.

Gamma matrix $\Gamma^{a \sp \alpha}_{\sp \sp \sp \beta}$ is
defined as follows,
\begin{eqnarray}
 \{ \Gamma^{a} ,\Gamma^{b} \} = 2 \eta ^{ab}.
\end{eqnarray}
We use the mostly plus metric; $\eta_{ab}\sim (-+...+)$.
We lower the spinorial indices
by charge conjugation matrix $C_{\alpha \beta }$.
\begin{eqnarray}
  \bar{\psi}_{\beta} = \psi ^{\alpha }C_{\alpha \beta }, \nonumber \\
  \Gamma^a _{\sp \sp \alpha \beta}=C_{\alpha \gamma }
  \Gamma^{a \sp \gamma}_{\sp \sp \sp \beta} .
\end{eqnarray}
$\Gamma^{a_1..a_n}_{\sp \sp \sp \sp \alpha\beta}(n=1,2,5,6,9,10)$ are 
symmetric matrices and $\Gamma^{a_1..a_n}_{\sp \sp \sp \sp \alpha\beta}
(n=0,3,4,7,8,11)$ are antisymmetric matrices.


\end{document}